\documentclass[reprint,showpacs,showkeys,prd,balancelastpage,nofootinbib]{revtex4}
\usepackage[utf8]{inputenc}
\usepackage{color}
\usepackage{amsmath}
\usepackage{amssymb}
\usepackage{graphicx}
\usepackage[unicode=true,
 bookmarks=false,
 breaklinks=false,pdfborder={0 0 1},backref=section,colorlinks=true]{hyperref}
\usepackage{amsfonts}
\usepackage{mdframed}
\usepackage{footnote}
\usepackage{graphicx}
\usepackage{float}
\usepackage[font=footnotesize,it]{caption}
\usepackage{tikz}
\usepackage{tikz-3dplot}

\hypersetup{
 linkcolor=blue,urlcolor=blue,citecolor=red}
\makeatletter
\@ifundefined{textcolor}{}
{
 \definecolor{BLACK}{gray}{0}
 \definecolor{WHITE}{gray}{1}
 \definecolor{RED}{rgb}{1,0,0}
 \definecolor{GREEN}{rgb}{0,1,0}
 \definecolor{BLUE}{rgb}{0,0,1}
 \definecolor{CYAN}{cmyk}{1,0,0,0}
 \definecolor{MAGENTA}{cmyk}{0,1,0,0}
 \definecolor{YELLOW}{cmyk}{0,0,1,0}
}
\makeatother

\input{tcilatex}

\begin{document}

\title{Generalized Extended Momentum Operator}
\author{M. Izadparast }
\email{habib.mazhari@emu.edu.tr}
\author{S. Habib Mazharimousavi}
\email{mustafa.halilsoy@emu.edu.tr}
\affiliation{Department of Physics, Faculty of Arts and Sciences, Eastern Mediterranean
University, Famagusta, North Cyprus via Mersin 10, Turkey}
\date{\today }

\begin{abstract}
We study and generalize the momentum operator satisfying the extended
uncertainty principle relation (EUP). This generalized extended momentum
operator (GEMO) consists of an arbitrary auxiliary function of position
operator, $\mu \left( x\right) $, in such a combination that not only GEMO
satisfies the EUP relation but also it is Hermitian. Next, we apply the GEMO
to construct the generalized one-dimensional Schr\"{o}dinger equation. Upon
using the so called point canonical transformation (PCT), we transform the
generalized Schr\"{o}dinger equation from $x$-space to $z$-space where in
terms of the transformed coordinate, $z$, it is of the standard form of the
Schr\"{o}dinger equation. In continuation, we study two illustrative
examples and solve the corresponding equations analytically to find the
energy spectrum.
\end{abstract}

\maketitle

\section{Introduction}

The idea of generalized momentum operator in quantum mechanics stems in the
Heisenberg uncertainty principle (HUP) \citep{1,2}. Let's recall that, any
simultaneous measurements of incompatible physical quantities in quantum
theory, such as position and momentum or time and energy, admit uncertainty.
For the case of the position and momentum operators where the HUP relation
states 
\begin{equation}
\Delta x\Delta p\geq \frac{1}{2}\left\vert \left\langle \left[ x,p\right]
\right\rangle \right\vert  \label{eq:1}
\end{equation}%
the more precisely one measures the position of a particle the less
information in momentum one gains. In order to modify the HUP, the minimum
length is proposed in quantum gravity \citep{3}, string theory \citep{4} and
non-commutative spacetimes due to quantum field in quantized spacetimes %
\citep{5,6}. Correspondingly, the generalization of the uncertainty
principle for minimum position and momentum has been presented in the form
of the extended and generalized uncertainty principle (EGUP) in \citep{1,2}
which is given by

\begin{equation}
\Delta x\Delta p\geq \frac{1}{2}\left( 1+\alpha \Delta x^{2}+\beta \Delta
p^{2}\right) .  \label{eq:2}
\end{equation}%
The implementation of quantum mechanics in gravity limits the measurement by
a minimal length in which $\beta $ is related to the Planck's length %
\citep{2} in accordance with the generalized uncertainty principle (GUP). In
the identical manner, the extended uncertainty principle (EUP) discusses the
cutoff in the minimum momentum in the (anti) de sitter spacetime with the
term $\alpha $ representing the radius of the curved spacetime \citep{7}. In %
\citep{8,9}, upon using an extended momentum operator on the primordial
perturbation spectrum during the inflation caused by quantum fluctuations is
investigated. The extended form of the momentum operator is obtained on the
(anti) de Sitter spacetime considering the deformation of the spacetime in %
\citep{10}. The authors in the latter research, by taking the gravitational
interactions into account and ignoring the curved spacetime, have also found
the corrections to the temperature of black holes. In \citep{11}, the
corrections of energy spectra for harmonic oscillator and Hydrogen atom is
acquired inserting the extended form of the momentum operator in the de
Sitter background. Again in \citep{12}, the quantum theory is investigated
for the (Anti) de Sitter spacetime in which the special case of EUP is
applied. Correspondingly, the deformed exponential and trigonometric
functions are attained using the deformed-derivative and deformed-integral.
Furthermore, in \citep{13} the classical approach to the (Anti) de Sitter
background, having deformed-mechanics, is comprehensively discussed in one
and $d-$dimensions. The study on the corrections to the Bekenstein-Hawking
entropy in the Schwarzschild black holes has been done in \citep{14} for the
EUP and GUP cases with considerable outcomes. Interestingly, in \citep{15}
the exact solutions of the Klein-Gordon and Dirac equations in the deformed
space for one-dimensional harmonic oscillator are obtained. Additionally, in %
\citep{16} the impact of EUP on the thermodynamics of the system at
high-temperature has been determined. In \citep{16} in the context of EUP,
the exact solutions of some $d-$dimensional potentials including an infinite
box, a harmonic and pseudo-harmonic oscillators have been obtained. In this
respect, the minimum momentum uncertainty imposed by the corresponding EUP,
due to the (Anti) de Sitter spacetime, has been found. The Ramsauer-Townsend
effect has been studied in the curved spacetime with corresponding EUP
correction in \citep{17}. In the latter work authors considered the WKB
approximation for one-dimensional Schr\"{o}dinger equation upon utilizing
various potentials. In \citep{18}, the effects of EUP and GUP on the
entropy-area relation on the apparent horizon of the FRW have been studied.
Costa Filho, et al. in \citep{19} showed that the translation operator
relates the metric of spacetime to the generalized momentum. This,
ultimately, gives the deformed Hamiltonian as a consequence of curved
spacetime. Upon considering the EUP corrections very recently Hamil in %
\citep{20} obtained the one dimensional harmonic oscillator in AdS and dS
background. Recently, Costa Filho et al. in \citep{21} have found
connections between the quantum harmonic oscillator in a deform space - due
to its EUP corrections - and the Morse potential in the regular space.

We aim, in the current research, to introduce a generalized extended
momentum operator (GEMO), in the sense that: i) it satisfies the generalized
EUP relation, as well as ii) it represents a physical quantity upon
satisfying the standard Hermiticity condition. A limited form of such a
generalization has been already considered in \citep{22}, where its
non-Hermitian version has been discussed in \citep{23}.

This paper is organized as follows. In Sec. II we introduce the generalized
EUP together with the GEMO. In Sec. III we apply the GEMO to the
one-dimensional Schr\"{o}dinger equation, and upon using PCT its transformed
version is obtained. Two explicit examples are studied in this section and
the paper is summarized in the Conclusion.

\section{Generalized momentum operator}

In the so called extended uncertainty principle (EUP) the commutation
relation of position and momentum operators is generalized to be 
\begin{equation}
\left[ x,p\right] =i\hbar \left( 1+\mu \left( x\right) \right)  \label{eq:4}
\end{equation}%
in which $\mu \left( x\right) $ is a real well-defined function of position
operator $x$. With the standard definition of the momentum operator, $%
p=-i\hbar \frac{d}{dx}$, Eq. (\ref{eq:4}) becomes simply $\left[ x,p\right]
=i\hbar $ with $\mu \left( x\right) =0$ while for the more interesting case
where $\mu \left( x\right) =\alpha x^{2}$ it has been found that %
\citep{8,9,13,15,16,17,20} 
\begin{equation}
p=-i\hbar \left( 1+\alpha x^{2}\right) \frac{d}{dx}.  \label{eq:5}
\end{equation}%
In finding (\ref{eq:5}), the only requirement was 
\begin{equation}
\left[ x,p\right] =i\hbar \left( 1+\alpha x^{2}\right)  \label{eq:6}
\end{equation}%
however adding any function of the position operator i.e., $x$, to the
obtained momentum operator mathematically does not change the commutation
relation (\ref{eq:6}). Hence, a general momentum operator satisfying (\ref%
{eq:6}) may be written as 
\begin{equation}
p=-i\hbar \left( 1+\alpha x^{2}\right) \frac{d}{dx}+f\left( x\right)
\label{eq:7}
\end{equation}%
in which $f\left( x\right) $ is a well-defined function of position operator 
$x$. To specify $f\left( x\right) $ we impose the Hermiticity condition on $%
p $ which is expected to be held by any physical quantity. This, in turn,
results in 
\begin{equation}
f\left( x\right) =-i\hbar x  \label{eq:8}
\end{equation}%
and consequently the Hermitian counterpart of the extended momentum operator
(\ref{eq:5}) becomes 
\begin{equation}
p=-i\hbar \left( 1+\alpha x^{2}\right) \frac{d}{dx}-i\hbar x.  \label{eq:9}
\end{equation}

Next, we generalize the above result in terms of EUP corresponding to the
commutation relation expressed in (\ref{eq:4}) by proposing the Hermitian
GEMO defined by 
\begin{equation}
p=-i\hbar \left( 1+\mu \left( x\right) \right) \frac{d}{dx}-i\hslash \frac{%
d\mu \left( x\right) }{2dx}.  \label{eq:10}
\end{equation}

\section{The generalized Schr\"{o}dinger equation}

Employing the generalized extended momentum operator (\ref{eq:10}), the
corresponding generalized Schr\"{o}dinger equation for a one-dimensional
quantum particle becomes 
\begin{equation}
i\hbar \frac{\partial }{\partial t}\psi \left( x,t\right) =\left( \frac{p^{2}%
}{2m}+V\left( x\right) \right) \psi \left( x,t\right) ,  \label{eq:11}
\end{equation}%
which upon considering $V\left( x\right) $ a time-independent potential and $%
\psi \left( x,t\right) =e^{-iEt/\hslash }\phi \left( x\right) $ we find the
time-independent Schr\"{o}dinger equation in its explicit form given by 
\begin{equation}
\frac{-\hbar ^{2}}{2m}\left( \left( 1+\mu \right) ^{2}\frac{d^{2}}{dx^{2}}%
+2\left( 1+\mu \right) \mu ^{\prime }\frac{d}{dx}+\frac{1}{2}\left( 1+\mu
\right) \mu ^{\prime \prime }+\frac{1}{4}\left( \mu ^{\prime }\right)
^{2}\right) \phi \left( x\right) +V\left( x\right) \phi \left( x\right)
=E\phi \left( x\right) .  \label{eq:12}
\end{equation}%
Here, $E$ is the conserved energy of the particle and a prime stands for the
derivative with respect to $x.$ The generalized Schr\"{o}dinger equation (%
\ref{eq:12}) can be transformed into a more familiar shape if we apply the
so-called \textquotedblleft
Point-Canonical-Transformation\textquotedblright\ (PCT), defined by 
\begin{equation}
\phi \left( x\right) =\frac{1}{\sqrt{1+\mu \left( x\right) }}\chi \left(
z\left( x\right) \right)   \label{eq:13}
\end{equation}%
and 
\begin{equation}
z=z\left( x\right) =\int^{x}\frac{1}{1+\mu \left( y\right) }dy+z_{0}
\label{eq:14}
\end{equation}%
in which $z_{0}$ is an integration constant. Upon applying (\ref{eq:13}) and
(\ref{eq:14}), the transformed generalized Schr\"{o}dinger equation becomes 
\begin{equation}
\frac{-\hbar ^{2}}{2m}\frac{d^{2}}{dz^{2}}\chi \left( z\right) +V\left(
x\left( z\right) \right) \chi \left( z\right) =E\chi \left( z\right) 
\label{eq:15}
\end{equation}%
which is in the form of the standard one-dimensional Schr\"{o}dinger
equation in the $z-$space.

\subsection{Illustrative example 1: $\protect\mu \left( x\right) =\protect%
\alpha ^{2}x^{2}$}

\begin{figure}[tbp]
\includegraphics[width=80mm,scale=0.7]{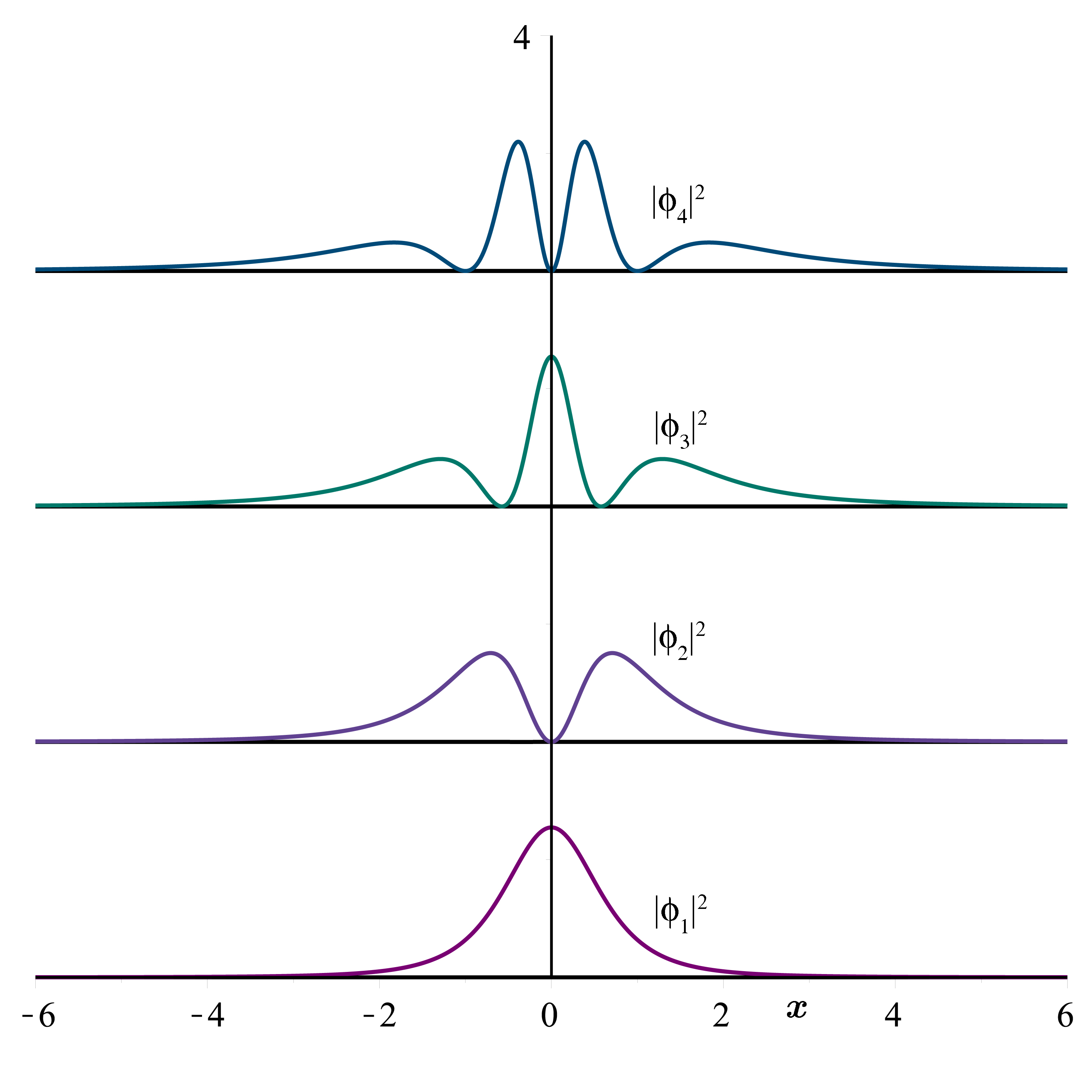}
\caption{The plot of $\left\vert \protect\phi _{n}\left( x\right)
\right\vert ^{2}$ in terms of $x$, Eq. (\protect\ref{eq:21}), for $n=1,2,3$
and $4$ with $\protect\alpha =1$. The free particle is confined within its
own generalized momentum such that the particle is localized around $x=0$
where its deviation from the standard momentum remains small.}
\label{Fig01}
\end{figure}

In this section we study a quantum particle whose generalized extended
momentum operator is defined by Eq. (\ref{eq:10}) with $\mu \left( x\right)
=\alpha ^{2}x^{2}$ in which $\alpha $ is a real constant of dimension $%
L^{-1}.$ Applying the PCT (\ref{eq:14}) we obtain 
\begin{equation}
z\left( x\right) =\frac{1}{\alpha }\arctan \left( \alpha x\right)
\label{eq:16}
\end{equation}%
or inversely 
\begin{equation}
x\left( z\right) =\frac{1}{\alpha }\tan \left( \alpha z\right)  \label{eq:17}
\end{equation}%
where we set the integration constant $z_{0}$ to be zero. We note that, for $%
x\in \mathbb{R}$, the transformed coordinate $z$ is confined i.e., $z\in %
\left[ -\frac{\pi }{2\alpha },\frac{\pi }{2\alpha }\right] .$ For a
zero-potential configuration in $x-$space, i.e., $V\left( x\right) =0$, one
finds the corresponding particle in an infinite well in the $z-$space where 
\begin{equation}
V\left( z\right) =\left\{ 
\begin{array}{cc}
0 & \left\vert z\right\vert \leq \frac{\pi }{2\alpha } \\ 
\infty & elsewhere%
\end{array}%
\right. .  \label{eq:18}
\end{equation}%
The eigenfunctions and eigenvalues of the reference equation (\ref{eq:15})
are given by 
\begin{equation}
\chi _{n}\left( z\right) =\left\{ 
\begin{array}{cc}
\sqrt{\frac{2\alpha }{\pi }}\cos \left( n\alpha z\right) , & odd-n \\ 
\sqrt{\frac{2\alpha }{\pi }}\sin \left( n\alpha z\right) , & even-n%
\end{array}%
\right.  \label{eq:19}
\end{equation}%
and 
\begin{equation}
E_{n}=\frac{n^{2}\hbar ^{2}\alpha ^{2}}{2m}  \label{eq:20}
\end{equation}%
respectively, where $n=1,2,3,...$. Furthermore, while the energy spectrum of
the target Schr\"{o}dinger equation, (\ref{eq:12}), is the same as (\ref%
{eq:20}) the normalized eigenfunctions, after (\ref{eq:13}), are given by 
\begin{equation}
\phi _{n}\left( x\right) =\left\{ 
\begin{array}{cc}
\sqrt{\frac{2\alpha }{\pi }}\frac{\cos \left( n\arctan \left( \alpha
x\right) \right) }{\sqrt{1+\alpha ^{2}x^{2}}}, & odd-n \\ 
\sqrt{\frac{2\alpha }{\pi }}\frac{\sin \left( n\arctan \left( \alpha
x\right) \right) }{\sqrt{1+\alpha ^{2}x^{2}}}, & even-n%
\end{array}%
\right. .  \label{eq:21}
\end{equation}%
In Fig. \ref{Fig01} we plot the first four states corresponding to the
quantum numbers $n=1,2,3$ and $4$ for $\alpha =1$. Unlike a free particle
with standard momentum operator which is equally likely to be found
everywhere on the position axis, here the extended momentum operator
confines the particle to be localized around smaller $x$.

It is remarkable to observe that, due to the zero potential i.e.,$V\left(
x\right) =0,$ the generalized momentum operator $p$ and the Hamiltonian, $H=%
\frac{p^{2}}{2m},$ commute. This implies that $\phi _{n}\left( x\right) $
are the simultaneous eigenstates of the Hamiltonian and the momentum with
eigenvalues $E_{n}$ and $\pm \sqrt{2mE_{n}},$ respectively. The extended
form of the momentum operator, i.e., 
\begin{equation}
p=-i\hbar \left( 1+\alpha ^{2}x^{2}\right) \frac{d}{dx}-i\hbar \alpha ^{2}x
\label{eq:22}
\end{equation}%
leads to a modified EUP where we expect a minimum uncertainty for the
momentum operator to be observed. This can be seen from 
\begin{equation}
\delta x\delta p\geq \frac{1}{2}\left\vert \left\langle \left[ x,p\right]
\right\rangle \right\vert  \label{eq:23}
\end{equation}%
where $\left[ x,p\right] =i\hbar \left( 1+\alpha ^{2}x^{2}\right) .$ The
latter implies 
\begin{equation}
\delta x\delta p\geq \frac{\hbar }{2}\left( 1+\alpha ^{2}\left\langle
x^{2}\right\rangle \right)  \label{eq:24}
\end{equation}%
which after knowing $\left( \delta x\right) ^{2}=\left\langle
x^{2}\right\rangle -\left\langle x\right\rangle ^{2}$, one finds 
\begin{equation}
\delta x\delta P\geq \frac{\hbar }{2}\left( 1+\alpha ^{2}\left( \left(
\delta x\right) ^{2}-\left\langle x\right\rangle ^{2}\right) \right) \geq 
\frac{\hbar }{2}\left( 1+\alpha ^{2}\left( \left( \delta x\right)
^{2}\right) \right) .  \label{eq:25}
\end{equation}%
Finally the momentum uncertainty is found to satisfy 
\begin{equation}
\delta p\geq \frac{\frac{\hbar }{2}\left( 1+\alpha ^{2}\left( \left( \delta
x\right) ^{2}\right) \right) }{\delta x}  \label{eq:26}
\end{equation}%
which obviously admits a minimum value for $\delta p$ at $\delta x=\frac{1}{%
\alpha }$, given by 
\begin{equation}
\left( \delta p\right) _{min}=\hbar \alpha .  \label{eq:27}
\end{equation}%
Using the eigenfunctions of the system, discussed above, Eq. (\ref{21}),
here we shall find $\delta x$ and $\delta p$ explicitly in order to
investigate the minimum value of the momentum uncertainty (\ref{eq:27}).
From the definition, 
\begin{equation}
\left( \delta p\right) ^{2}=\left\langle p^{2}\right\rangle -\left\langle
p\right\rangle ^{2}  \label{eq:28}
\end{equation}%
where 
\begin{equation}
\left\langle p^{2}\right\rangle =\left\langle \phi _{n}\left\vert
p^{2}\right\vert \phi _{n}\right\rangle =2mE_{n}  \label{eq:29}
\end{equation}%
and 
\begin{equation}
\left\langle p\right\rangle =\left\langle \phi _{n}\left\vert p\right\vert
\phi _{n}\right\rangle =0  \label{eq:30}
\end{equation}%
which amount to $\delta p=n\hbar \alpha $. Recall that, $n=1,2,3,...$, the
minimum uncertainty for the momentum operator occurs in the ground state
where $n=1$ and $\left( \delta p\right) _{min}=\hbar \alpha $, in agreement
with (\ref{eq:27}). On the other hand, since $\left\langle x\right\rangle =0$
and $\left\langle x^{2}\right\rangle =\frac{2n-1}{\alpha ^{2}},$ we obtain $%
\delta x=\frac{\sqrt{2n-1}}{\alpha }$. Finally, one gets 
\begin{equation}
\delta x\delta p\geq n\sqrt{2n-1}\hbar  \label{eq:31}
\end{equation}%
which clearly satisfies (\ref{eq:24}).

\subsection{Illustrative example 2: $\protect\mu \left( x\right) =e^{-%
\protect\gamma x}-1$}

\begin{figure}[tbp]
\includegraphics[width=80mm,scale=0.7]{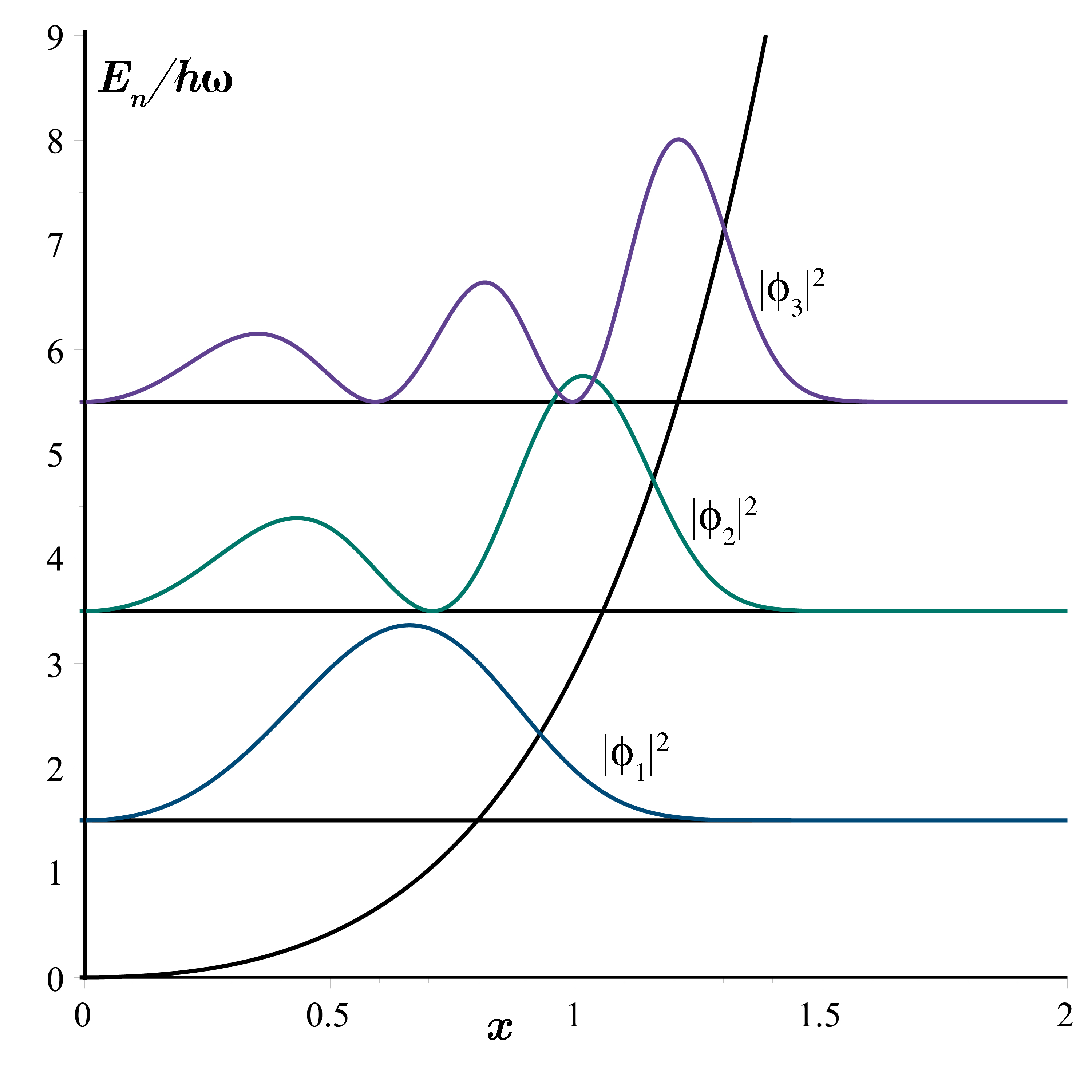}
\caption{Plot of $\left\vert \protect\phi _{n}\left( x\right) \right\vert
^{2}$ (Eq. (\protect\ref{eq:40})) in terms of $x$ for $n=0,1$ and $2$ with $%
V_{0}=1,m=1,\hbar =1$ and $\protect\alpha =1.$ The potential (\protect\ref%
{eq:35}) is also plotted together with the energy of each state on the
vertical line. We observe that in higher levels the probability densities
moves away from the infinite wall located at $x=0$.}
\label{Fig02}
\end{figure}

In this final section we study another important generalized extended
momentum operator with 
\begin{equation}
\mu \left( x\right) =e^{-\gamma x}-1  \label{eq:32}
\end{equation}%
in which $\gamma $ is a real constant. Similar to the previous example we
apply the PCT (\ref{eq:14}) which implies 
\begin{equation}
z=\frac{1}{\gamma }\left( e^{\gamma x}-1\right)  \label{eq:33}
\end{equation}%
and 
\begin{equation}
\phi \left( x\right) =e^{\gamma x/2}\chi \left( z\right)  \label{eq:34}
\end{equation}%
where we set the integration constant to be $-\frac{1}{\gamma }.$
Furthermore, we assume a modified-half-Morse type potential in the $x-$%
space, i.e., 
\begin{equation}
V\left( x\right) =\left\{ 
\begin{array}{cc}
\infty & x\leq 0 \\ 
V_{0}\left( 1-e^{\gamma x}\right) ^{2} & x>0%
\end{array}%
\right.  \label{eq:35}
\end{equation}%
in which $V_{0}>0.$ In $z-$space the potential becomes 
\begin{equation}
V\left( z\right) =\left\{ 
\begin{array}{cc}
\infty & z\leq 0 \\ 
\frac{1}{2}m\omega ^{2}z^{2} & z>0%
\end{array}%
\right.  \label{eq:36}
\end{equation}%
where $\omega ^{2}=\frac{2\gamma ^{2}V_{0}}{m}$. The standard solution of
the quantum-half SHO is available in any text book given by $E_{n}=\hbar
\omega \left( 2n+\frac{3}{2}\right) $ and 
\begin{equation}
\chi _{n}\left( z\right) =\frac{1}{\sqrt{2^{n-1}n!}}\left( \frac{m\omega }{%
\pi \hslash }\right) ^{1/4}e^{-m\omega z^{2}/2\hbar }H_{n}\left( \sqrt{\frac{%
m\omega }{\hbar }}z\right) ,  \label{eq:37}
\end{equation}%
with Hermite polynomials defined by 
\begin{equation}
H_{n}\left( t\right) =\left( -1\right) ^{n}e^{t^{2}}\frac{d^{n}}{dt^{n}}%
\left( e^{-t^{2}}\right) .  \label{eq:38}
\end{equation}%
Herein, the quantum number $n=0,1,2,3,4,...$. Finally the eigenvalues and
eigenfunctions of the original system in $x-$space are given by 
\begin{equation}
E_{n}=\hbar \omega \left( 2n+\frac{3}{2}\right)  \label{eq:39}
\end{equation}%
and 
\begin{equation}
\phi _{n}\left( x\right) =\frac{1}{\sqrt{2^{n-1}n!}}\left( \frac{\gamma 
\sqrt{2mV_{0}}}{\pi \hslash }\right) ^{1/4}\exp \left[ -\frac{1}{2\gamma }%
\left( \frac{\sqrt{2mV_{0}}}{\hbar }\left( e^{\gamma x}-1\right) ^{2}-\gamma
^{2}x\right) \right] H_{n}\left( \sqrt{\frac{\sqrt{2mV_{0}}}{\hbar \gamma }}%
\left( e^{\gamma x}-1\right) \right)  \label{eq:40}
\end{equation}%
respectively. In Fig. \ref{Fig02} we plot the probability density $%
\left\vert \phi _{n}\left( x\right) \right\vert ^{2}$ in terms of $x$ for
the first three states with $V_{0}=1,m=1,\hbar =1$and $\alpha =1.$ It is
observed that for higher level states, the probability density moves away
from the infinite wall.

\section{Conclusion}

We introduced the GEMO in terms of an auxiliary function of position
operator, $\mu \left( x\right) $. The generalization refers to the fact that
this modified momentum operator is Hermitian as well as satisfying the EUP
commutation relation. Using this momentum operator we, then, constructed the
one-dimensional Schr\"{o}dinger equation in terms of $\mu \left( x\right) $.
Applying the PCT, we transformed the explicit form of the resulting Schr\"{o}%
dinger equation, from $x-$space to $z-$space where it became simpler and
familiar in terms of its similarity to the standard one-dimensional Schr\"{o}%
dinger equation. Following that we solved the Schr\"{o}dinger equation for
two different settings of $\mu \left( x\right) .$ In the first case we
considered $\mu \left( x\right) =\alpha ^{2}x^{2}$ with a zero potential $%
V\left( x\right) =0$ and in the second case we set $\mu \left( x\right)
=e^{-\gamma x}-1$ and $V\left( x\right) $ given in (\ref{eq:35}). In both
illustrative examples we found exact solutions for the energy eigenstates
and eigenvalues. Moreover, in the first case we have shown that the momentum
admits a minimum uncertainty which was expected. This was shown from the EUP
commutation relation as well as explicitly from the energy eigenstates.
Finally we add that, a two or three dimensional generalization could be
naturally considered which we leave it open for further work.

\end{document}